# Scalable single-microring hybrid III-V/Si lasers for emerging narrow-linewidth applications

Jiawei Wang,[1] Xiang Li,[2] Xin Guo,[1] Ter-Hoe Loh,[3] Luigi Ranno,[4] Chongyang Liu,[5] Rusli,[1] Hong Wang,[1] and Jia Xu Brian Sia[1,4,*]

[1]*School of Electrical and Electronic Engineering, Nanyang Technological University, 50 Nanyang Avenue, 639798, Singapore*
[2]*Shenzhen Pinghu Laboratory, 2 Laolang Road, Shenzhen 518111, China*
[3]*Fingate Technologies, 8 Cleantech Loop, 637145, Singapore*
[4]*Department of Material Science & Engineering, Massachusetts Institute of Technology, Cambridge, M.A., USA*
[5]*Temasek Laboratories, Nanyang Technological University, 50 Nanyang Avenue, 637553, Singapore*
*\*jiaxubrian.sia@ntu.edu.sg*

**Abstract:** Silicon photonics, compatible with large-scale silicon manufacturing, is a disruptive photonic platform that has indicated significant implications in industry and research areas (e.g., quantum, neuromorphic computing, LiDAR). Cutting-edge applications such as high-capacity coherent optical communication and heterodyne LiDAR have escalated the demand for integrated narrow-linewidth laser sources. To that effect, this work seeks to address this requirement through the development of a high-performance hybrid III-V/silicon laser. The developed integrated laser, utilizes a single microring resonator (MRR), demonstrating single-mode operation with a side mode suppression ratio (SMSR) exceeding 40 dB, with laser output power as high as 16.4 mW. Moving away from current hybrid/heterogeneous laser architectures that necessitate multiple complex control, the developed laser architecture requires only two control parameters. Importantly, this serves to streamline industrial adoption by reducing the complexity involved in characterizing these lasers, at-scale. Through the succinct structure and control framework, a narrow laser linewidth of 2.79 kHz and low relative intensity noise (RIN) of -135 dB/Hz are achieved. Furthermore, optical data transmission at 12.5 Gb/s is demonstrated where a signal-to-noise ratio (SNR) of 10 dB is measured.

## 1. Introduction

As an integrated optical technology predicated upon large-scale silicon manufacturing, silicon photonics (SiP) has proved to be disruptive [1-5]. Thus far, the field has witnessed significant commercialization, building upon cornerstone development in the area [6]. Furthermore, new avenues of application are opening up, where instances include neuromorphic computing [7], LIDAR [8], environmental sensing [9], and healthcare diagnostics [10].

Among emerging photonics applications, one of the foremost requirements is the development of integrated narrow-linewidth laser sources [11-14]. The acquisition of NeoPhotonics Inc., by Lumentum Inc. for 918 million USD in 2021 is indicative of the current market demand, and the application for these laser technologies [15]. Instances include coherent optical communications, where laser linewidth influences modulation complexity, receiver sensitivity and digital signal processing [16,17]. Lasers with ultra-narrow linewidth are mandated in current advanced coherent optical systems to maximize capacity across greater distances. On the other hand, narrow-linewidth lasers are imperative in heterodyne LIDAR systems to maintain a stable frequency reference over entire measurement intervals [18]. This essential temporal coherence enables precise measurements of depth, velocity, optical phase shifts, and resilience to noise. It can be concluded that the development of narrow-linewidth lasers is of great technological importance.

To that effect, significant efforts have been directed towards linewidth narrowing in lasers [13,19]. While highly-efficient, electrically-pumped monolithic lasers remain elusive in SiP, the platform emerged as the ideal candidate to achieve the above. This is a corollary of the advanced silicon manufacturing techniques and processes, where low waveguide propagation losses can be attained. Through the integration of low-loss passive SiP laser cavities with III-V amplifiers, the photon lifetime in hybrid/heterogeneous lasers can be significantly increased in comparison to their III-V counterparts. The majority of current demonstrations have involved the hybrid/heterogeneous integration of the III-V gain section, and the implementation of two or three microring resonators (MRRs) in the passive Si/SiN laser cavity [20-24]. The assembly of MRRs produces a Vernier spectrum, allowing spectral control of the Vernier peak to a specific longitudinal mode via the respective thermo-optic control of each MRR. This facilitates the wavelength-tuning functionality of these lasers [25,26]. However, the control of these lasers features a large parameter space where complex tuning is necessitated. For instance, in the passive SiP cavity, one will have to consider the thermo-optic heater voltage control of the laser longitudinal mode, as well as the Vernier spectrum borne from two or more MRRs. This is in addition to current injection levels of the III-V gain section. While superior performance has been demonstrated via these laser architectures, large-scale characterization, as required for commercialization, will be highly time-consuming. Therefore, reducing the complexity of the SiP laser cavity, while still indicating high phase purity, is imperative.

In this work, by leveraging on the III-V/silicon hybrid platform, the abovementioned is addressed through the following. 1.) Via the design of the laser longitudinal mode spacing and the 3-dB bandwidth of a single MRR resonator, combined with the implementation of thermo-optic phase shifters, single-mode operation (side mode suppression ratio (SMSR) > 40 dB) is demonstrated. Laser output power as high as 16.4 mW is measured. 2.) The laser architecture demonstrated in this work only requires two control parameters (Fig. 1: laser current, MRR heater), indicating a huge contrast from lasers which requires 4 [21,22] or 5 [23,24] control values. This eases potential industrial adoption by reducing the laser characterization and control complexity. 3.) Critically for narrow laser linewidth applications, this work indicates that narrow laser linewidth (linewidth = 2.79 kHz) can be attained with just a single MRR within the passive laser cavity. Additionally, the laser indicates low relative intensity noise (RIN) of -135 dB/Hz. Finally, data transmission experiments indicating opened eyes at 12.5 Gb/s is demonstrated.

## 2. Laser concept

### 2.1 Laser operating principle

The 3D schematic diagram of the single MRR hybrid III-V/Si laser diode is illustrated in Fig. 1 with XYZ coordinates, where the inset shows the 2D top-view microscope image with XY coordinates. The III-V gain section facilitates electrically-pumped optical emission and gain, while the SiP laser cavity enables the filtering of a single longitudinal mode for single-mode laser operation. At the laser output facet, the III-V waveguide is angled to the normal, to increase laser output power, where the reflectivity at the facet is ~27 % (mirror 1). Antireflection coating is implemented on the III-V coupling facet to the SiP laser cavity, and the III-V waveguide is angled at 6° to suppress gain ripples, which would impact single-mode laser operation. The reflectivity of the facet is measured to be 0.005%, where coupling between the III-V and SiP is enabled by a silicon nitride (SiN) mode interposer angled at 11.39°, determined via Snell's law. A SiN/Si spot-size converter (SSC), shown at the inset of Fig. 1, is utilized to transit the optical mode from the SiN layer, through a 300-nm $SiO_2$ spacing in between, down to the Si waveguide layer where the passive silicon laser cavity is implemented. The optical mode is filtered by the drop spectrum of the MRR (detailed in section 2.2) and is then reflected via a Sagnac loop mirror reflector with 100 % reflectivity (mirror 2) [27]. Oscillation essential for laser operation is enabled between mirror 1 and 2.

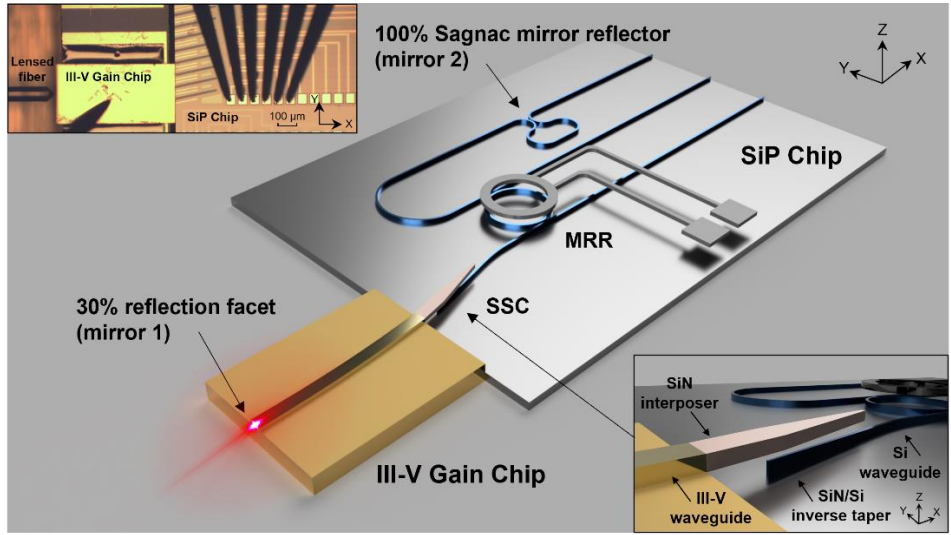

Fig. 1. 3D schematic of the edge-coupled hybrid III-V/silicon laser diode, where the assembly consists of a III-V gain section and a silicon passive laser cavity. Inset shows the 2D top-view microscope image of the laser, illustration of the SiN interposer, and SiN/Si spot-size converter. The XYZ coordinates of the laser are indicated.

Specifically, the operating principle of the single MRR hybrid III-V/silicon laser diode is elucidated in Fig. 2(a). The overlap between the MRR drop spectrum and the specific laser longitudinal mode experiences the least loss in the laser cavity, leading to single-mode lasing via mode competition. Two aspects of the laser design are crucial for achieving single-mode laser operation. Firstly, the power coupling coefficient ($\kappa^2$) between the MRR and the bus waveguide is tailored such that the 3-dB bandwidth of the MRR drop spectrum is smaller than twice the laser longitudinal mode spacing. This ensures that the desired laser longitudinal mode exhibits a high modal transmittance difference (MTD) compared to other modes in this ring resonance. With increment in laser bias currents ($I_{bias}$), the gain of the III-V section will increase, and the longitudinal mode subject to the lowest cavity losses will have the propensity to lase. However, the alignment between the laser longitudinal modes and MRR drop spectrum peak may shift due to $I_{bias}$ variations. To address this, thermo-optic phase shifters are mounted on top of the MRR, where voltages ($V_{mrr}$) are applied to supplement longitudinal mode alignment. Through the precise alignment of the desired laser longitudinal mode with the peak of the MRR drop spectrum at different $I_{bias}$ values, the laser longitudinal mode will experience the lowest loss near the center of the amplified spontaneous emission (ASE) gain spectrum, thus single-mode lasing operation is achieved at each different $I_{bias}$. As a result, single-mode laser operation is realized with one single MRR, requiring only two control parameters ($I_{bias}$ and $V_{mrr}$), which largely simplifies the device architecture. The laser performance will be detailed in section 3.1 of this paper.

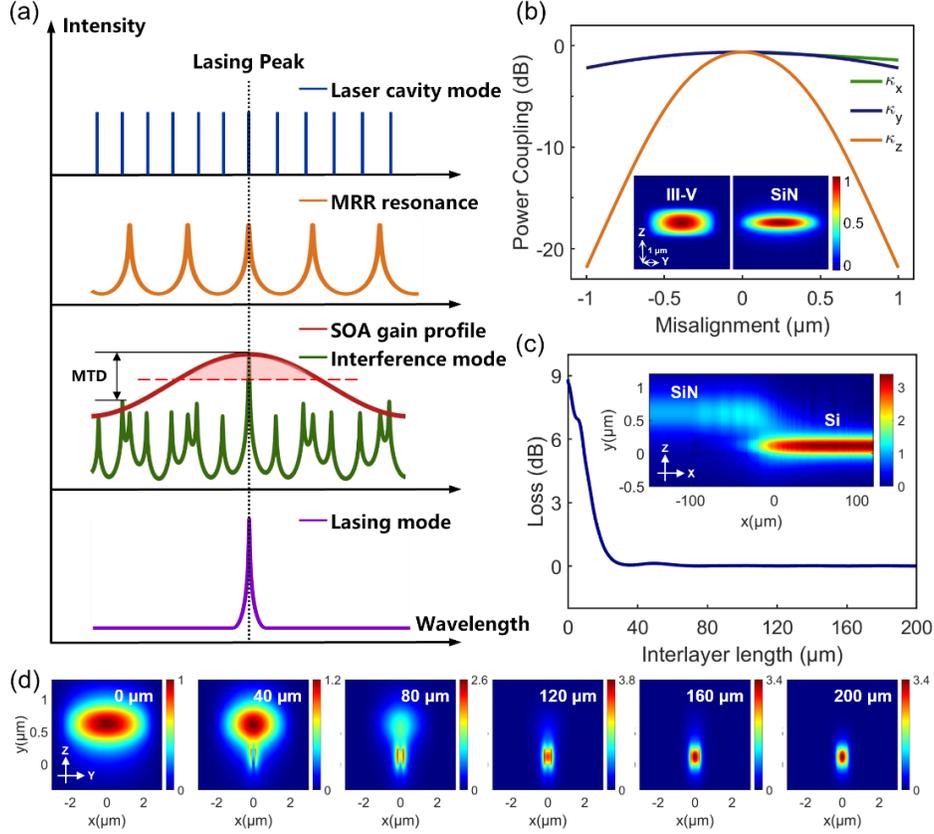

Fig. 2. (a) Operating principle of the single MRR III-V/Si hybrid laser diode. (b) Alignment tolerance of the III-V and SiN waveguide interposer in the x, y, and z axis. Inset shows the mode profiles of the III-V and SiN waveguide. (c) SiN/Si SSC length optimization using EME calculations. Inset shows the electric field distribution (XZ-plane) as the waveguide mode transits along a 200 µm-long converter. (d) Evolution of cross-sectional electric field distribution (YZ-plane) at different length (0, 40, 80, 120, 160, 200 µm) as the waveguide mode transits from the SiN interposer to the Si waveguide along a 200 µm-long SSC.

## *2.2 Photonic design*

The optical design leverages the commercial 220 nm SiP platform design reported by Sia et al. [28]. The $1/e^2$ near-field mode field diameter (MFD) of the III-V waveguide is $4.0 \times 1.2$ µm$^2$ (YZ-plane). In order to facilitate efficient coupling between the III-V section and the passive silicon cavity, the SiN layer is utilized, comprising of a SiN waveguide interposer with a cross-section of $5 \times 0.4$ µm$^2$ and a $1/e^2$ near-field MFD of $5.0 \times 1.2$ µm$^2$ (YZ-plane). At the optimal coupling condition, an optical power overlap of 86% (-0.66 dB) can be achieved between the III-V waveguide and SiN waveguide interposer. The alignment tolerance in the XYZ-direction between the III-V and the SiN waveguide is analyzed in Fig. 2(b), with mode profiles at the III-V and SiN interposer facets shown in the insets. The desired alignment tolerance is indicated on the X/Y-axis, where misalignments of ±0.5 µm only correspond to excess losses of 0.98/1.04 dB, respectively. The Z-axis, on the other hand, displays lower tolerance, where misalignments of ±0.5 µm increase excess loss by 6.35 dB. This is inherently limited by the size of the optical mode at the III-V facet in the Z-axis (1.2 µm). To improve coupling tolerance in the X/Y/Z-directions, methods that increase the near-field MFD at the III-V facet can be implemented. Instances of such approaches, while not exhaustive, include the design of tapered gain regions

[29]. Subsequently, SiN or Si inverse tapers can be implemented on the passive SiP laser cavity to enhance alignment tolerances.

Following the coupling of the optical emission from the III-V gain section, the optical mode is transferred from the SiN device layer to the Si layer via a SiN/Si spot-size converter (SSC), based on a push-pull taper structure [30]. The C-band fundamental waveguides are determined to be 450 nm wide. To be conservative, the width of taper tips at the SiN and Si layers are chosen to be 300 and 150 nm, respectively, such that the features can be reliably defined through fabrication. The length of SiN-Si SSC is optimized using the Eigenmode Expansion method, shown in Fig. 2(c), where it is determined that for lengths exceeding 30 µm, low-loss adiabatic conversion between the SiN and Si waveguide layers can be attained. The length of the SiN/Si SSC is selected to be 200 µm. The cross-sectional electric field distribution in the XZ-plane, and its evolution along the YZ-plane as the waveguide mode transits from the SiN interposer to the Si waveguide along the 200 µm-long SSC are shown at the inset of Fig. 2(c), and Fig. 2(d). These simulations indicate an efficient mode transition from SiN to Si layer.

**Table 1. Effective Optical Path Length of Laser Components**

| Laser component | Gain section | SiN interposer | SiN/Si SSC | Si waveguides | MRR | Total |
|---|---|---|---|---|---|---|
| Optical Length/mm | 3.336 | 0.256 | 0.388 | 1.255 | 12.726 | 17.573 |

The effective optical length of the hybrid laser is tailored specifically for laser single-mode operation. Table 1 presents a breakdown of the effective optical path length in terms of each constituent component: gain section, SiN interposer, SiN/Si SSC, Si waveguides, and MRR. This gives rise to a computed longitudinal mode spacing of 0.068 nm. With that, the design requirement of the MRR is considered. The $\kappa^2$ between the MRR and bus waveguide is shown as a function of the coupling gap. The spectra at the drop port of the MRR when both coupling gaps are symmetrical are given by [31]:

$$P_{drop} = \left| \frac{-\kappa^* \kappa \alpha_{1/2mrr} e^{j\theta_{1/2mrr}}}{1 - t^* t \alpha_{mrr} e^{j\theta_{mrr}}} \right|^2 \tag{1}$$

Where $\kappa$ refers to the electric field coupling coefficient between the MRR and the bus waveguide ($\kappa = \sqrt{\kappa^2}$). $t$ refers to the electric field transmission where $t^2 = 1 - \kappa^2$. $\alpha_{1/2mrr}$ and $\theta_{1/2mrr}$ refers to the loss and phase change incurred after the lightwave propagates half a round in the MRR, respectively; from one coupling section to another coupling section. Similarly, $\alpha_{mrr}$ and $\theta_{mrr}$ refers to the losses and phase change after the propagation of one full trip in the MRR. Fig. 3(a) indicates the theoretical 3-dB bandwidth of MRR drop spectra subject to $\kappa^2$. In order to ensure that the 3-dB bandwidth of the drop spectra is smaller than twice the laser longitudinal mode spacing (2 × 0.068 nm), an MRR to bus waveguide coupling gap of 230 nm is selected, corresponding to a $\kappa^2$ of 0.039 and a 3-dB bandwidth of 0.06 nm.

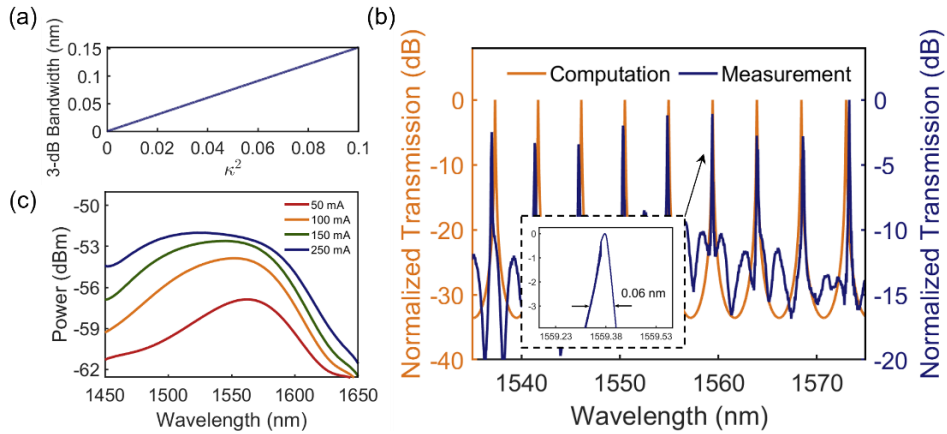

Fig. 3. (a) The 3-dB bandwidth of MRR drop spectra as a function of $\kappa^2$. (b) The computed and measured drop port spectra of the MRR. Inset: measured MRR drop peak at 1559.38 nm. (c) The smoothed ASE curve of the gain section at different $I_{bias}$.

In Fig. 3(b), the computed and measured drop port spectra of MRR are shown, where significant agreement is observed between both spectra. The measured drop spectra indicate a 3-dB bandwidth of 0.06 nm and an FSR of 4.5 nm. In Fig. 3(c), the smoothed ASE curves of the gain section at various $I_{bias}$ values are presented. At lower $I_{bias}$ when the laser is about to lase, the peak of the ASE spectrum is located around 1563 nm, which becomes the target working wavelength for the designed single-mode laser. As abovementioned, the design requirement mandates the MRR 3-dB bandwidth to be smaller than twice the laser longitudinal mode spacing, which is measured to be 0.057 nm under pre-lasing conditions. These measurements affirm that the laser diode has been fabricated close to the design.

## 3. Experimental data

### 3.1 Laser characterization

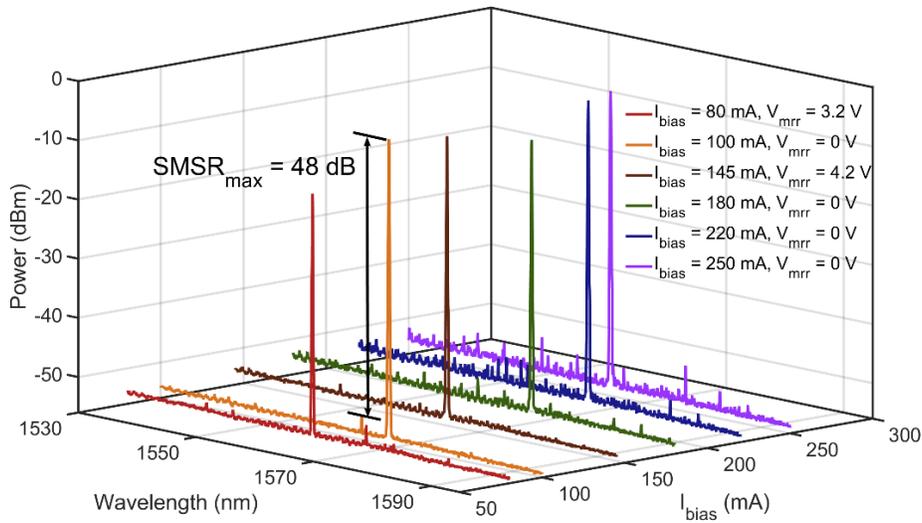

Fig. 4. Laser evolution spectrum at different $I_{bias}$ and $V_{mrr}$ settings.

The III-V gain section (InP-based, multi-quantum well) and the SiP laser cavity, placed on two separate 6-axis precision alignment stages, are edge coupled, thereby forming a hybrid III-V/Si laser cavity (Fig. 1). Through independent testing structures, the III-V to Si edge coupling losses are measured to be 3.9 dB. The excess losses arise primarily from misalignments, particularly in the Z-axis (Fig. 2(b)).

To characterize the laser spectrum, the output emission is collected via a lensed fiber, linked to an optical spectrum analyzer (Yokogawa AQ6375). In Fig. 4, we present the optical spectrum at various $I_{bias}$ and $V_{mrr}$ settings. Single-mode laser operation with an SMSR in excess of 40 dB, reaching as high as 48 dB, is observed, validating the laser design. As the laser comprises only two control parameters ($I_{bias}$, $V_{mrr}$), tuning of the laser becomes straightforward. At each $I_{bias}$ setting value, a $V_{mrr}$ applied to the MRR is used to adjust the overlap between the MRR drop spectrum and the laser longitudinal mode. Herein, we refer to the overlap mode as the interference mode. Once the interference mode peak with the lowest loss aligns with the peak of ASE gain spectrum, it experiences the least loss and thus lases via mode competition as elucidated above. At some $I_{bias}$ values, the interference mode peak is located at the center of ASE gain spectrum, thus no excess voltage is required ($V_{mrr} = 0$ V). By actively tuning $V_{mrr}$ at each different $I_{bias}$ value, a single-mode lasing state can be achieved with high power and high SMSR across the entire $I_{bias}$ range.

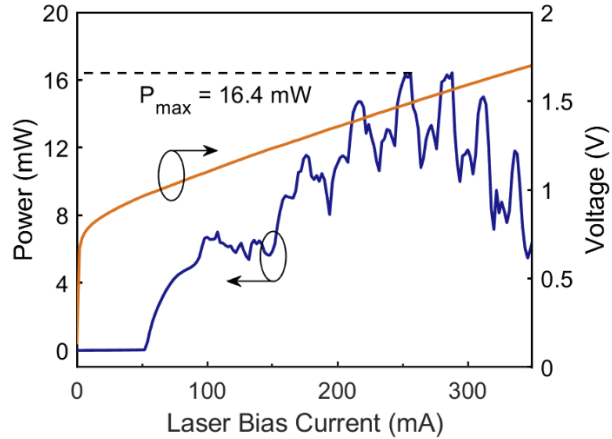

Fig. 5. LIV curve of the single MRR III-V/Si hybrid laser ($V_{mrr} = 0$).

Following, the output power of the laser as a function of $I_{bias}$ (L-I) is characterized by coupling the laser emission into a free-space power detector when $V_{mrr} = 0$ (as shown in Fig. 5); for reference, the I-V curve of the laser diode is also presented. Room temperature continuous wave operation is realized with a low threshold current of 52 mA. The output power is demonstrated to reach as high as 16.4 mW at $I_{bias} = 256$ mA, beyond which it is limited by the two-photon absorption of silicon [32]. Mode hopping is observed due to multi-mode lasing, but it can be mitigated by the active control of $V_{mrr}$ along with increasing $I_{bias}$ (Fig. 4) [11]. Thermal control is imperative for III-V-based laser diodes. A notable advantage of hybrid III-V/silicon integration over their heterogeneous counterparts is the ability to implement a thermally conductive and reactive path at the III-V section [33,34]. To specifically address the issue of laser thermal dissipation, the III-V gain section is implemented P-side down, mounted to an AuSn submount. Placing the quantum wells, and consequently the heat source, closer to the P-side of the gain section facilitates effective heat dissipation.

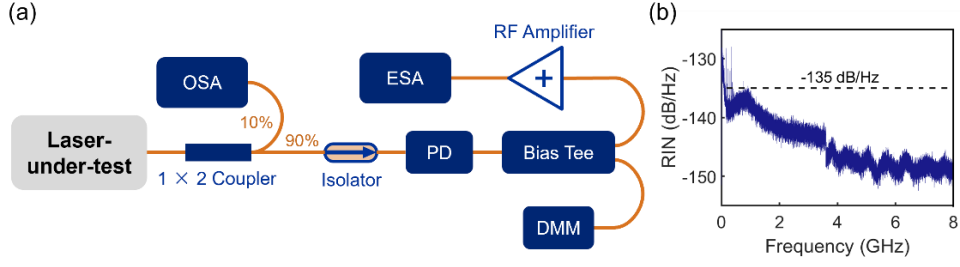

Fig. 6. (a) Experimental setup for measuring RIN. (b) The measured RIN of the single MRR III-V/Si hybrid laser ($I_{bias}$ = 180 mA, $V_{mrr}$ = 0 V).

$$RIN(f) = \frac{((N_{rf}(f) - N_{th}(f))/G(f)}{P_{elec}} \qquad (2)$$

The relative intensity noise (RIN) of the laser is measured using the setup illustrated in Fig. 6(a). First, the laser emission is split through a $1 \times 2$ coupler (10/90 %), where 10 % of the laser power is directed to the optical spectrum analyzer (OSA) to monitor laser wavelength and single-mode stability. The remaining laser output power is then sent to an optical isolator, followed by a high-speed photodetector (PD). A bias tee is utilized to separate the PD output current into its direct current (DC) and alternating current (AC) components. The DC signal represents the detected photocurrent, $I_0$ (corresponding to the electrical power $P_{elec}$), which is measured with a digital multimeter (DMM). Meanwhile, the AC signal is amplified through an RF amplifier ($G(f)$ = 28 dB, $f$ = 0.1 – 18 GHz), connected to an electrical spectrum analyzer (ESA). The measured ESA output constitutes the total noise term $N_{rf}$, comprising laser RIN as well as thermal noise, $N_{th}$. A laser RIN of -135 dB/Hz is determined via equation (2) when $I_{bias}$ = 180 mA and $V_{mrr}$ = 0 V [35]. From a general photonic system perspective, low RIN is mandated for a wide range of applications. Low laser intensity fluctuations enable small signals to be resolved, maintaining signal integrity, and optimizing the signal-to-noise ratio (SNR) in a plethora of photonic systems [36,37].

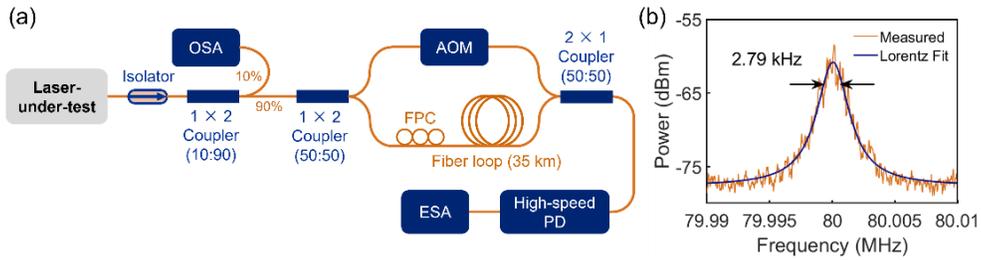

Fig. 7. (a) Experimental setup for measuring laser linewidth. (b) Narrow linewidth of 2.79 kHz is measured with Lorentz curve fitting method for the single MRR III-V/Si hybrid laser ($I_{bias}$ = 100 mA, $V_{mrr}$ = 0 V).

The measurement setup utilized to characterize the laser linewidth is illustrated in Fig. 7(a), where a 35 km-long fiber loop is integrated into the conventional self-heterodyne measurement approach. Laser emission is coupled into an optical isolator through a lensed fiber, then a $1 \times 2$ coupler (10/90 %). Similar to the RIN measurement demonstrated above, 10 % of the laser power is routed to an OSA to monitor the wavelength and single-mode stability of the laser. The remaining 90 % of the laser output is linked to another $1 \times 2$ coupler (50/50 %), where one of the output ports is connected to an acousto-optic modulator (AOM), and the other to the 35 km long fiber, following the fiber-polarization controller (FPC). Subsequently, the two ports are combined via a 2 × 1 coupler (50/50 %), and finally directed to a high-speed PD and ESA. The AOM operates at 80 MHz, inducing a frequency shift where the beat signal is expected to occur. During linewidth measurement, the ESA is averaged 5 times.

The laser linewidth is measured at $I_{bias}$ = 100 mA and $V_{mrr}$ = 0 V. The beat signal at 80 MHz is zoomed in and shown in Fig. 7(b), indicating a full-width half-maximum (FWHM) of 2.79 kHz, assuming a Lorentzian line shape. The narrow linewidth measured demonstrates the phase purity of the single-mode laser, implying high compatibility with prospective applications (e.g., long-reach and large-bandwidth optical communications [38,39], coherent LIDAR [40,41]) that require ultra-low phase noise.

### 3.2 Optical data transmission experiment

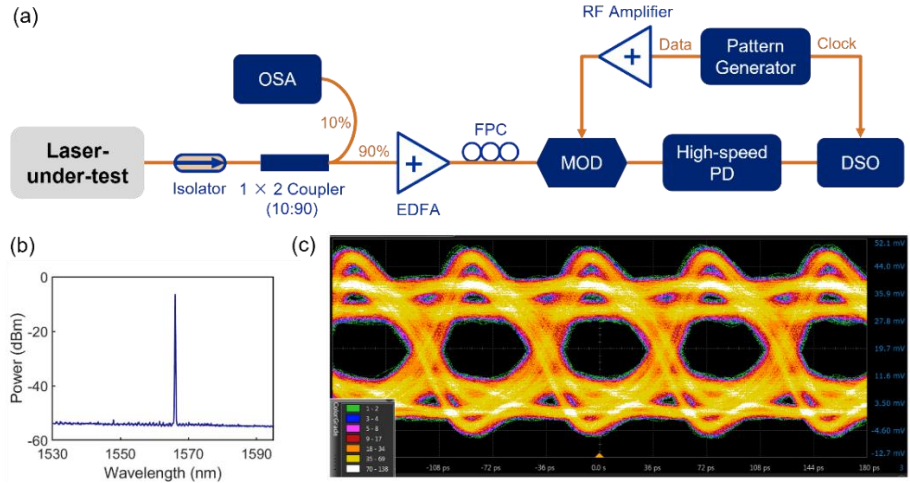

Fig. 8. (a) Experimental setup for the optical data transmission experiment. (b) Corresponding laser spectrum for the optical data transmission experiment ($I_{bias}$ = 145 mA, $V_{mrr}$ = 4.17 V). (c) Measured eye diagram with a 12.5 Gb/s PRBS source.

In this preceding section, the single MRR hybrid III-V/Si laser is utilized as a source for a data transmission experiment; the experiment setup is illustrated in Fig. 8(a). The corresponding laser spectrum implemented ($I_{bias}$ = 145 mA and $V_{mrr}$ = 4.17 V) is shown in Fig. 8(b), where λ = 1566 nm. A bulk lithium niobate modulator is used to modulate the laser output in the data transmission link. A 12.5 Gb/s pseudo-random binary sequence (PRBS) with a pattern length of $2^7$-1 is generated by a pattern generator (Anritsu MP1763C); the generated pattern is limited to 12.5 Gb/s. The PRBS produces a peak-to-peak output voltage ($V_{p-p}$) of 0.25 V, which is amplified to 6 $V_{p-p}$ and applied to the modulator. The emission of the hybrid III-V/Si laser (Fig. 8(b)) is coupled to the modulator, where the resultant output is directed to a high-speed PD and a 30 GHz real-time oscilloscope (DSO 93004L). Clear eye opening is observed as indicated in Fig. 8(c), where an SNR of 10 dB is measured.

### 4. Conclusion

In this work, a narrow linewidth III-V/silicon hybrid laser is demonstrated utilizing a streamlined architecture. Via careful design of the laser longitudinal mode spacing and the 3-dB bandwidth of a single MRR resonator, combined with the implementation of thermo-optic phase shifters, single-mode operation (SMSR > 40 dB) is achieved across the entire SOA bias current range. The measured laser output power reaches as high as 16.4 mW. Through the laser design, only two control parameters ($I_{bias}$ and $V_{mrr}$) are required to achieve single-mode lasing, thereby reducing the complexity of laser characterization and control. This eases potential industrial adoption. Narrow linewidth of 2.79 kHz and low RIN of -135 dB/Hz are measured, and clear opened eyes are obtained with an SNR of 10 dB in the optical data transmission experiment. The favorable optical performance of the III-V/silicon hybrid laser implies good potential in various prospective applications, including long-reach and large-bandwidth optical communications, as well as coherent LIDAR, where ultra-low phase noise is essential.

**Funding.** This work was supported by the Ministry of Education (MOE) Singapore under Grant MOE-T2EP50121-0005.

**Acknowledgments.** Jia Xu Brian Sia would like to acknowledge the Ministry of Education/NTU College of Engineering International Postdoctoral Fellowship.

**Disclosures.** The authors declare no conflicts of interest.

**Data availability.** Data underlying the results presented in this paper are not publicly available at this time but may be obtained from the authors upon reasonable request.